\documentclass[fleqn,twoside]{article}%
\usepackage{amsmath}
\usepackage{amsfonts}
\usepackage{amssymb}
\usepackage{graphicx}
\usepackage{cite}
\def\be{\begin{equation}}

\def\ee{\end{equation}}
\def\bes{\begin{equation}\begin{split}&}
\def\es{\end{split}}
\def\bi{\bibitem}

\begin{document}
\title{The role of cosmological constant in f(R, G) gravity.}
\author{Abhik Kumar Sanyal$^1$ and Chandramouli Sarkar $^2$}

\maketitle
\noindent
\begin{center}
\noindent
$^1$ Dept. of Physics, Jangipur College, Murshidabad, West Bengal, India - 742213. \\
$^2$17A, Lake Road, Kolkata, India- 700029.\\

\end{center}
\footnotetext{\noindent
Electronic address:\\
\noindent
$^1$ sanyal\_ ak@yahoo.com\\
$^2$ csarkar123@yahoo.co.in}

\noindent
\abstract{Einstein-Hilbert action is supplemented by Gauss-Bonnet squared term, its phase-space structure is constructed and canonical quantization is performed. Resolution of a contradiction that emerges in the process, requires the presence of other fields at least in the form of vacuum energy-density, commonly known as the cosmological constant. This reveals the very importance of the presence of other fields at least in the form of cosmological constant, in the very early universe.}\\

\noindent
keywords: Higher Order theory; Canonical Formulation.\\

\maketitle

\flushbottom

\section{Introduction}

General Theory of Relativity (GTR) or any of its modified versions treats all the fields on the same footing. Although all the fields are quantized much prior to the gravitational field - being described by geometry, it is not known which field should be treated as fundamental, say for the creation of the universe. It is well known that GTR is non-renormalizable and a renormalized theory has not been found as yet. However, all attempts to cast a renomalized theory of gravity (different versions of string theories) indicate that in the very universe the Einstein-Hilbert action must be supplemented by higher order curvature invariant terms. The existence of vacuum de-Sitter (anti-de-Sitter) solution, being a primary check for the choice of particular combination of higher-order terms, does not require the presence of fields of any kind other than gravity. Further, lot of research had been oriented \cite{PC1, PC2, PC3, PC4, PC5, PC6, PC7} regarding creation of particles from strong gravitational field. In this sense, it is a preconceived notion that geometry must be treated as fundamental. In this article we construct the phase-space structure of a recently advocated (to explain late-time cosmic evolution of the universe) geometric theory associated with Einstein-Hilbert action in the presence of Gauss-Bonnet squared term alone, by analysing constrained dynamics following Dirac's algorithm. In the process, we show that a quantized version of the theory, leads to some sort of inconsistency, viz: while classical inflationary solution admits a positive coupling parameter, extremization of the effective potential although admits inflationary solution, requires a negative coupling parameter. The problem is resolved under the addition of a cosmological constant in the said action. The fact that the cosmological constant is the sum of zero point vacuum energy density of all possible fields existing in the very early universe, the resolution of the said contradiction reveals that geometry might not be treated as fundamental. We also perform semiclassical approximation following the same technique adopted earlier\cite{HO1, HO2, HO3, HO4, HO5, HO6}.  \\

As mentioned, in the present manuscript we supplement Einstein-Hilbert action with Gauss-Bonnet squared term. Gauss-Bonnet term is topologically invariant in four dimension, as a result it doesn't contribute to the classical field equations. Therefore to have some contribution, it is usually coupled to a dilatonic scalar field, which arises naturally in the weak field limits of several effective string theories \cite{ST1, ST2, ST3, ST4, ST5, ST6, ST7}. Such an action is well suited to explain late-time cosmic evolution, particularly the recently observed accelerated expansion of the universe, after a long matter dominated Friedmann-like evolution $a \propto t^{2\over 3}$, where, $a$ is the scale factor. It also admits transient double crossing of the phantom divide line, which is not excluded from the recent observations. The question arises regarding the presence of the scalar field itself. While there is no indication of the presence of a scalar field in any of its form in the late universe, a dilatonic coupled Gauss-Bonnet interaction seems to be unrealistic. It is therefore reasonable to discard the scalar field and try to explain the late-time cosmic evolution otherwise. This is possible under modification of the left hand side of Einstein's equation, which corresponds to incorporating higher order curvature invariant terms in the Einstein-Hilbert action, which as discussed, is essential for a renormalized theory of gravity. One, out of such many attempts has been successfully made by taking higher powers of Gauss-Bonnet term in the action, which does not require coupling to a scalar field even in $4$-dimension and leads to GTR as the low energy limit of an affective quantum theory \cite{GB1}. Soon after, Neutron star solutions were presented \cite{GB2}, and also relativistic massive objects viz. compact star solutions and their dynamical stability were studied for a viable power law model of $f(G)$ \cite{GB3}. The model also has been proved to be successful in unifying early inflation and late-time cosmic evolution \cite{GB4, GB5, GB6, GB7}. It is therefore required to test the viability of such an action in the very early universe.\\

In this paper we therefore couple Einstein-Hilbert sector of the gravitational action with the Gauss-Bonnet squared term and study its viability in the very early universe in regard of the quantum dynamics and inflationary scenario. It is important to mention that while Gauss-Bonnet term leads to second order field equations, its higher powers truly behaves as higher order theory. Canonical formulation of such higher-order theory is clearly non-trivial. We organize the present manuscript as follows. In the following section, we take up Gauss-Bonnet squared term along with the linear gravitational sector, find the field equations in the background of isotropic and homogeneous space-time which have been found to admit vacuum de-Sitter solution. We study the behaviour of the action in the radiation dominated era and also the inflation. In section 3, we cast the action in canonical form and formulate the phase-space structure, following Dirac's algorithm. Thereafter we follow the standard canonical quantization scheme and attempted semiclassical approximation about the classical de-Sitter solution. In the process we observe that the theory suffers from certain pathologies. The situation changes dramatically under the influence of a cosmological constant, which we study next in section 4. We conclude in section 5.

\section{The action, field equations, classical solutions and Inflation.}

We start with the following $f(R, \mathcal{G})$ action expressed in its simplest form as,
\be \label{A} A = \int \left[\alpha R + \gamma \mathcal{G}^2 \right]\sqrt{-g} d^4 x,\ee
where, $\alpha = {1\over 16\pi G}$, and $\gamma$ are coupling constants, while the Gauss-Bonnet term is $\mathcal{G} = R^2 - 4 R_{\mu\nu}R^{\mu\nu} + R_{\alpha\beta\mu\nu}R^{\alpha\beta\mu\nu}$. In homogeneous and isotropic Robertson-Walker metric

\be ds^2 = -N(t)^2 dt^2 + a(t)^2 \left[ {dr^2\over 1-kr^2} + r^2 d\theta^2 - r^2 \sin^2{\theta} d\phi^2\right],\ee
under the choice $z= a^2$,

\be\begin{split}& R = \frac{6}{N^2}\left(\frac{\ddot a}{a}+\frac{\dot a^2}{a^2}+N^2\frac{k}{a^2}-\frac{\dot N\dot a}{N a}\right) = {6\over N^2}\left[{\ddot z\over 2z} + N^2 {k\over z} - {1\over 2}{\dot N\dot z\over N z}\right]\\&
\mathcal{G} = {24\over N^3 a^3}(N\ddot a - \dot N\dot a)\Big({\dot a^2\over N^2} + k\Big)= {12\over N^2}\left({\ddot z\over z} - {\dot z^2\over 2 z^2} -{\dot N\dot z\over N z}\right)\left({\dot z^2\over 4N^2 z^2} + {k\over z}\right).\end{split}\ee
The action therefore reads as

\be \label{Action1} A = \int \left[ {6\alpha\over N^2}\left[{\ddot z\over 2z} + N^2 {k\over z} - {1\over 2}{\dot N\dot z\over N z}\right] +{144\gamma\over N^4}\left({\ddot z\over z} - {\dot z^2\over 2 z^2} -{\dot N\dot z\over N z}\right)^2\left({\dot z^2\over 4N^2 z^2} + {k\over z}\right)^2 \right]Nz^{3\over 2} dt \int d^3 x.\ee
Field equations ($k = 0$) under the inclusion of a barotropic fluid with pressure ($p$) and energy density $\rho$ read as,

\bes \label{FE1}2\alpha\left({\ddot z\over z}-{\dot z^2\over 4z^2}\right)+12\gamma\left[{\dot z^4\ddddot z\over z^5}+{8\dot z^3\ddot z\dddot z\over z^5}-{9\dot z^5\dddot z\over z^6}+{6\dot z^2\ddot z^3\over z^5}-{135 \dot z^4\ddot z^2 \over 4 z^6}+{159 \dot z^6\ddot z\over 4z^7} - {195 \dot z^8\over 16 z^8}\right] = - p,\end{split}\ee

\be \label{FE2} {3\alpha \dot z^2\over 2z^2}+18\gamma\left[{\dot z^5 \dddot z\over z^6} + {3\dot z^4 \ddot z^2\over 2z^6} - {9 \dot z^6\ddot z\over 2 z^{7}} + {15 \dot z^8\over 8 z^8}\right] = \rho,\ee
together with the energy conservation equation,

\be \label{EE} \dot \rho + 3{\dot a\over a} (\rho + p) = \dot \rho + {3\dot z\over 2 z} (\rho + p) = 0.\ee

\subsection{Classical solution:}

One can clearly notice that the The above field equations do not admit power law solution in any form in the radiation dominated era ($p = {1\over 3}\rho, ~ \rho z^2 = \rho_{ro}, ~ \rho_{ro}$ being a constant which measures the mount of radiation left in the present day universe). In the literature, it is found that people are mostly interested in unifying early inflation with ate time cosmic acceleration. The behaviour of a theory in the radiation dominated era is not usually explored, Here, we observe that, despite the claim that the theory can explain late-time cosmic evolution with appropriate accelerated expansion, it confronts with the standard model in the radiation dominated era. This tells upon the structure formation and the CMBR we observe. Therefore the theory in no way is viable to explain the evolution history of the universe. This is the first pathology appearing in connection with the theory under investigation.\\

However, in the present manuscript, we are indeed interested in very early vacuum dominated ($p = 0 = \rho$) era. The above set of field equations admit the following vacuum de-Sitter inflationary solution in the form

\be\label{Sol2} z = z_0 e^{2\lambda t} \Rightarrow a = a_0 e^{\lambda t}; ~~\mathrm{where,}~~\lambda = \frac{1}{2}\left(\frac{2\alpha}{3\gamma}\right)^{\frac{1}{6}} = \frac{1}{2}\left(\frac{M_p^2}{3\gamma}\right)^{\frac{1}{6}},\ee
where, $M_P$ is the Planck's mass. We shall require the above vacuum de-Sitter solution to compute on-shell semicalssical approximation.

\subsection{Inflation:}

Let us first describe the background evolution by a set of horizon flow functions (the behaviour of Hubble distance during
inflation) starting from
\be \epsilon_0 = {d_H\over d_{Hi}},\ee
where, $d_H = {H^{-1}}$ is the Hubble distance, or more commonly the horizon, in the unit $c = 1$. Suffix $i$ stands for the era when inflation was initiated. Let us now define a hierarchy of functions in a systematic way by,
\be \epsilon_{m + 1} = {d\ln|\epsilon_m|\over dN},~~~~m\ge 0, ~~~ N = \ln{a\over a_i} = \int H dt,\ee
where $H$ is the Hubble parameter. one can compute $\epsilon_1 = {d ln dH\over dN}$, which is the logarithmic change of Hubble distance per e-fold expansion N. It is in fact the first slow-roll parameter $\epsilon_1 = \dot{d}_H = − {\dot H\over H^2}$ implying that the Hubble parameter almost remains constant during inflation. The above hierarchy allows one to compute $\epsilon_2= {d \ln \epsilon_1\over dN} = {1\over H}{\dot \epsilon_1\over \epsilon_1}$, leading to $\epsilon_1\epsilon_2 = d_H \ddot d_H = -{1\over H^2}\left({\ddot H\over H} -  2{\dot H^2\over H^2}\right)$. In the same manner higher slow roll parameters may be computed. Equation \eqref{FE2} may now be expressed as,

\be \alpha + 96 \gamma \left[2H^6 \left\{1 + {1\over H^2}\left({\ddot H \over H}-2{\dot H^2\over H^2}\right)\right\}  + 7 H^6 \left(1 + {\dot H\over H^2}\right)^2 - 8 H^6\left(1 + {\dot H\over H^2}\right) - 2 H^6\right]= 0,\ee
or in terms of the slow-roll parameters,

\be \alpha + 96\gamma\left[2H^6(1 - \epsilon_1\epsilon_2) + 7 H^6(1 - \epsilon_1)^2 - 8 H^6(1-\epsilon_1) - 2H^6\right] = 0.\ee
Now, under slow roll approximation $\epsilon_m << 0$, one arrives at,

\be\alpha = 96 H^6\gamma,\ee which is solved to yield

\be \label{Inf} a(t) = a_0 \exp {\left[\left(\alpha\over 96\gamma\right)^{1\over 6} t\right]}.\ee
This slow roll inflationary solution resembles exactly with the one obtained in view of an effective Lagrangian $L = R + \beta R^2 + \gamma R^4$, in the regime when $96\gamma H^4 >> 6 \beta$, i.e. in the regime when $R^2$ term is neglected. This implies that Gauss-Bonnet squared term $(\mathcal{G}^2)$ effectively plays the same role as $R^4$ term in the modified theory of gravity, at least in the background of homogeneous and isotropic Robertson-Walker space-time. \cite{GB7}. It is important to mention for later reference that both the classical de-Sitter solution \eqref{Sol2} and the slow-roll inflationary solution \eqref{Inf} are admissible for a positive coupling parameter $\gamma$.

\section{Canonical formulation and quantization}

Performing some algebraic manipulation and integrating the appropriate terms appearing in the action \eqref{Action1} by parts, following total derivative terms are found

\be \Sigma = \int \left[{6\alpha\sqrt z\dot z\over 2N} - 24^2 \gamma\left({\dot z^7\over 448 N^7 z^{11\over 2}}+{k\dot z^5\over 40 N^5 z^{9\over 2}}+{k^2\dot z^3\over 12 N^3 z^{7\over 2}} \right)\right]d^3 x\ee
which vanishes identically in the Dirac's formalism where $\delta h_{ij}|_{\partial\mathcal{V}} = 0 = \delta K_{ij}|_{\partial\mathcal{V}}$, where, $h_{ij}$ is the induced three-metric and $K_{ij}$ is the extrinsic curvature tensor. Therefore the action finally takes the form,

\be \label{A1} \begin{split} A = & \int\Bigg[6\alpha\left(-{\dot z^2\over 4 N \sqrt z}+ k N\sqrt z\right) + 24^2\gamma\Bigg\{\left({\dot z^4\over 64 N^7 z^{9\over 2}}+{k\dot z^2\over 8 N^5 z^{7\over 2}}+{k^2\over 4 N^3 z^{5\over 2}}\right)\ddot z^2\\&
-\left({\dot N \dot z^5\over 32 N^8 z^{9\over 2}} + {k\dot N\dot z^3\over 4 N^6 z^{7\over 2}} + {k^2\dot N\dot z\over 2 N^4 z^{5\over 2}}\right)\ddot z
-{15\dot z^8\over 1792 N^7 z^{13\over 2}}-{13 k\dot z^6\over 160N^5 z^{11\over 2}}-{11 k^2\dot z^4\over 48 N^3 z^{9\over 2}}\\&+{\dot N^2\dot z^6\over 64 N^9 z^{9\over 2}}+{k\dot N^2\dot z^4\over 8 N^7 z^{7\over 2}}+{k^2\dot N^2\dot z^2\over 4 N^5 z^{5\over 2}}\Bigg\} \Bigg]dt\int d^3x.\end{split}\ee
It should be mentioned that canonical formulation must be performed in terms of the pair of basic variables, which are $h_{ij},~K_{ij}$. In the present situation $h_{ij} = a^2 \delta_{ij} = z\delta_{ij}$ and $K_{ij} = -{\dot h_{ij}\over 2N} = - {a\dot a\over N} = -{\dot z\over 2N}$. This is the reason why we have expressed the action in terms of $z = a^2$ from the very beginning.

\subsection{Analysing the constraint:}

Now to follow Dirac's algorithm, we substitute $\dot z = N x$, i.e. $\ddot z = N\dot x + \dot N x$, and therefore can express the point Lagrangian in the form,

\be\begin{split} L = &~ 6\alpha\left(-{N x^2\over 4\sqrt z}+ k N\sqrt z\right) \\&
+ 288\gamma \left[\left({x^4\over 32 N z^{9\over 2}}+{k x^2\over 4Nz^{7\over 2}}+{k^2\over 2Nz^{5\over 2}}\right)\dot x^2
- {15 N x^8\over 896 z^{13\over 2}} - {13 k N x^6\over 80 z^{11\over 2}} - {11 k^2 N x^4\over 24 z^{9\over 2}}\right] + u\left({\dot z\over N} - x\right),\end{split}\ee
where we have treated the expression ${\dot z\over N} - \dot x$ as a constraint and introduced it through the Lagrange multiplier $u$ in the above point Lagrangian. The canonical momenta are,

\be\label{p1} p_x = {288\gamma\over N}\left({x^4\over 16 z^{9\over 2}}+{k x^2\over 2 z^{7\over 2}}+{k^2\over z^{5\over 2}}\right) \dot x; \;\;\;p_z = {u\over N};\;\;p_N = 0 = p_u.\ee
The primary Hamiltonian therefore reads as

\be\begin{split} H_{p1} = & {N p_x^2\over 576\gamma \left({x^4\over 16 z^{9\over 2}}+{k x^2\over 2 z^{7\over 2}}+{k^2\over z^{5\over 2}}\right)} + \dot z{u\over N} + 6\alpha\left({N x^2\over 4\sqrt z}- k N\sqrt z\right)\\& \hspace{3.8 cm}+
{36\gamma N x^4}\left({15 x^4\over 112 z^{13\over 2}} + {13 k x^2\over 10 z^{11\over 2}} + {11 k^2\over 3 z^{9\over 2}}\right) - u {\dot z\over N} + u x.
\end{split}\ee
Now introducing the constraints $\phi_1 = Np_x - u \approx 0$ and $\phi_2 = p_u \approx 0$ through the Lagrange multipliers $u_1$ and $u_2$ respectively, we get

\be\begin{split} H_{p1} = & N\Bigg[{z^{9\over 2}p_x^2\over 36\gamma(x^2 + 4 k z)^2} + 6\alpha\left({x^2\over 4\sqrt z}- k \sqrt z\right)+ {36\gamma x^4}\left({15 x^4\over 112 z^{13\over 2}} + {13 k x^2\over 10 z^{11\over 2}} + {11 k^2\over 3 z^{9\over 2}}\right)\Bigg]\\&
\hspace{3.8 cm} + u_1(Np_x - u) + u_2 p_{u} + u x.\end{split}\ee
Note that the Poisson brackets $\{x, p_x\} = \{z, p_z\} = \{u, p_u\} = 1$, hold. Now constraint should remain preserved in time, which are exhibited through the following Poisson brackets

\be \dot \phi_1 = \{\phi_1, H_{p1}\} = - u_2 - N{\partial H_{p1}\over \partial z} \approx 0 \Rightarrow u_2 = - N{\partial H_{p1}\over \partial z};\;\;\;\dot \phi_2 = \{\phi_2, H_{p1}\} = - (u_1+x) \approx 0\Rightarrow u_1 = x.\ee
Therefore the primary Hamiltonian is modified to

\be\begin{split} H_{p2} = & N\left[x p_z + { z^{9\over 2}p_x^2\over 36\gamma(x^2 + 4 k z)^2}+  6\alpha\left({x^2\over 4\sqrt z}- k \sqrt z\right)+{36\gamma x^4}\left({15 x^4\over 112 z^{13\over 2}} + {13 k x^2\over 10 z^{11\over 2}} + {11 k^2\over 3 z^{9\over 2}}\right)\right] - N p_u{\partial H_{p1}\over \partial z}.\end{split}\ee
As the constraint should remain preserved in time in the sense of Dirac, so

\be \dot \phi_1 = \{\phi_1, H_{p2}\} = N{\partial H_{p1}\over \partial z} - N\left[{\partial H_{p1}\over \partial z}-Np_u {\partial^2 H_{p1}\over \partial z^2}\right]\approx 0 \Rightarrow p_u = 0.\ee
Thus, finally the phase-space structure of the Hamiltonian, being free from constraints reads as

\be\label{H}\ H = N\left[x p_z + {z^{9\over 2}p_x^2\over 36\gamma(x^2 + 4 k z)^2} +\alpha\left({3 x^2\over 2\sqrt z}- 6k \sqrt z\right)+36\gamma x^4\left({15 x^4\over 112 z^{13\over 2}} + {13 k x^2\over 10 z^{11\over 2}} + {11 k^2\over 3 z^{9\over 2}}\right)\right] = N\mathcal{H}.\ee

\noindent
It is important to note that the Einstein-Hilbert sector appears only in the potential part in the above Hamiltonian. The action \eqref{A1} may now be expressed in canonical ADM form ($k = 0$) as,

\be A = \int\big(\dot z p_z + \dot x p_x  - \ N \mathcal{H}\big) dt d^3 x = \int\big(\dot {h}_{ij}p^{ij} + \dot {K}_{ij}\Pi^{ij} - \ N \mathcal{H}\big) dt d^3 x.\ee
In the above equation, $p^{ij}$ and $\Pi^{ij}$ are the momenta canonical to $h_{ij}$ and $K_{ij}$ respectively. The presence of $x p_z$ term in the Hamiltonian \eqref{H} clearly dictates that as in the cases of higher order theories studied earlier, it also leads to Schr\"odinger-like equation, and hence quantum mechanical probabilistic interpretation should be admissible. It is also required to check for the sake of consistency, if the same classical field equations \eqref{FE1} and \eqref{FE2} are arrived at from the above Hamiltonian \eqref{H}. To avoid complication, we express the Hamilton's equations \eqref{H} for $k = 0$, as,

\be\label{H1} \mathcal{H} = x p_z + {z^{9\over 2}p_x ^2\over 36 \gamma x^4} + {135\gamma x^8\over 28 z^{13\over 2}} + {3\alpha x^2\over 2\sqrt z}. \ee
Hamilton's equations are,

\bes \label{He1}\dot x = {z^{9\over 2}p_x\over 18\gamma x^4}; ~~~~~\dot z = x;
~~~~\dot p_x = -p_z +{z^{9\over 2}p_x^2\over 9\gamma x^5} -{270\gamma x^7\over 7 z^{13\over 2}} -{3\alpha x\over \sqrt z};\\&
\dot p_z = -{z^{7\over 2} p_x^2\over 8\gamma x^4} + {1755\gamma x^8\over 56 z^{15\over 2}} +{3\alpha x^2\over 4 z^{3\over 2}} =-18\gamma\left[{9\dot z^4 \ddot z^2\over 4 z^{11\over 2}}-{195\dot z^8\over 112 z^{15\over 2}}\right] +{3\alpha \dot z^2\over 4z^{3\over 2}}.\end{split}\ee
From the first relation of the above set of Hamilton's equations, we find,

\bes \label{He2} p_x = {18\gamma x^4 \dot x\over z^{9\over 2}} = {18\gamma \dot z^4 \ddot z\over z^{9\over 2}};~~\dot p_x  = 18\gamma \left[{\dot z^4\dddot z\over z^{9\over 2}} + {4 \dot z^3\ddot z^2\over z^{9\over 2}}- {9\dot z^5\ddot z\over 2 z^{11\over 2}}\right]\\&
\ddot p_x =18\gamma\left[{\dot z^4\ddddot z\over z^{9\over 2}}+{12\dot z^3\ddot z\dddot z\over z^{9\over 2}}-{9\dot z^5\dddot z\over z^{11\over 2}}+{12\dot z^2\ddot z^3\over z^{9\over 2}}-{81 \dot z^4\ddot z^2 \over 2 z^{11\over 2}}+{99 \dot z^6\ddot z\over 4 z^{13\over 2}}\right]\\&
\mathrm{and~ so,~}p_z = -18\gamma\left[{\dot z^4\dddot z\over z^{9\over 2} }+ {2\dot z^3\ddot z^2\over z^{9\over 2}} - {9\dot z^5\ddot z\over 2z^{11\over 2}} + {15\dot z^7\over 7 z^{13\over 2}}\right] - {3\alpha x\over \sqrt z}.\end{split}\ee
Now, derivative of the second relation of the first equation yields,

\bes \label{He3}\ddot p_x = 18\gamma\left[{4\dot z^3 \ddot z\dddot z\over z^{9\over 2}} + {6\dot z^2\ddot z^3\over z^{9\over 2}} -{27\dot z^4 \ddot z^2\over 4 z^{11\over 2}} - {15\dot z^6\ddot z\over z^{13\over 2}}+ {195\dot z^8\over 16 z^{15\over 2}}\right] - {3\alpha \ddot z\over \sqrt z} + {3\alpha \dot z^2\over 4z^{3\over 2}}.\end{split}\ee
Equating relation $\ddot p_x$ between relations \eqref{He2} and \eqref{He3}, we obtain,

\bes \label{a}\alpha\left({\ddot z\over z}-{\dot z^2\over 4z^2}\right) + 6\gamma\left[{\dot z^4\ddddot z\over z^5}+{8\dot z^3\ddot z\dddot z\over z^5}-{9\dot z^5\dddot z\over z^6}+{6\dot z^2\ddot z^3\over z^5}-{135 \dot z^4\ddot z^2 \over 4 z^6}+{159 \dot z^6\ddot z\over 4z^7} - {195 \dot z^8\over 16 z^8}\right] = 0,\end{split}\ee
This is equation \eqref{FE1}. From the other set, one can also find the $(^0_0)$ equation of Einstein. However, the simplest way is to express the Hamiltonian in terms of configuration space variables and then setting it to zero, to find

\be \label{00} E = -18\gamma\left[{\dot z^5 \dddot z\over z^{9\over 2}} + {3\dot z^4 \ddot z^2\over 2z^{9\over 2}} - {9 \dot z^6\ddot z\over 2 z^{11\over 2}} + {15 \dot z^8\over 8 z^{13\over 2}}\right] - {3\alpha \dot z^2\over 2\sqrt z} = 0.\ee
This is the ($^0_0$) component \eqref{FE2} of the field equations. For a cross check, we take time derivative of the energy equation to find

\bes -{\dot E\over 3\sqrt z\dot z} = 6\gamma\left({\dot z^4\ddddot z\over z^5}+{8\dot z^3\ddot z\dddot z\over z^5}-{9\dot z^5\dddot z\over z^6}+{6\dot z^2\ddot z^3\over z^5}-{135 \dot z^4\ddot z^2 \over 4 z^6}+{159 \dot z^6\ddot z\over 4z^7} - {195 \dot z^8\over 16 z^8}\right) + \alpha\left({\ddot z\over z}-{\dot z^2\over 4z^2}\right) = 0,\end{split}\ee
which is equation \eqref{a}. Thus, we have cross checked, and there is no inconsistency at the classical level.

\subsection{Canonical quantization:}

Now canonical quantization leads to

\be i\hbar x{\partial \Psi\over \partial z} = { z^{9\over 2}\over 36\gamma}\widehat{{p_x^2\over (x^2 + 4 k z)^2}}\Psi
+\left[ \alpha\left({3x^2\over 2\sqrt z}- 6k \sqrt z\right)+{36\gamma x^4}\left({15 x^4\over 112 z^{13\over 2}} + {13 k x^2\over 10 z^{11\over 2}} + {11 k^2\over 3 z^{9\over 2}}\right)\right]\Psi.\ee
Operator ordering is required in the first term of the right hand side. Performing Weyl operator ordering, ($n$ being the operator ordering index) one obtains

\be \label{Q1} i\hbar z^{-{9\over 2}}{\partial \Psi\over \partial z} = -{\hbar^2\over 36\gamma x(x^2 + 4kz)^2}\left({\partial^2\over \partial x^2} + {n\over x}{\partial\over \partial x}\right)\Psi + \left[{3\alpha (x^2 - 4k z)\over 2 x z^5} + 36\gamma x^3\left({15 x^4\over 112 z^{11}} + {13 kx^2\over 10 z^{10}} + {11 k^2\over 3 z^4} \right) \right]\Psi.\ee
Finally, under a change of variable $\sigma = z^{11\over 2}$, so that $z^{-{9\over 2}}{\partial\over \partial z} = {11\over 2}{\partial\over \partial \sigma}$ one obtains

\bes \label{Q2} i\hbar{\partial\Psi\over \partial\sigma} = -{\hbar^2\over 198\gamma x(x^2 + 4k \sigma^{2\over 11})^2}\left({\partial^2\over \partial x^2} + {n\over x}{\partial\over \partial x}\right)\Psi + V_e(x,\sigma)\Psi\\& \mathrm{where,}~~V_e(x,\sigma) =
\left[{3\alpha\over 11 x} \left({x^2\over \sigma^{10\over 11}}- {4k\over \sigma^{8\over 11}}\right) + {72\gamma x^3\over 11}\left({15 x^4\over 112 \sigma^2} + {13 kx^2\over 10 \sigma^{20\over 11}} + {11 k^2\over 3 \sigma^{8\over 11}} \right) \right],\end{split}\ee
is the effective potential. In the above Schr\"odinger-like equation, $\sigma = a^{11}$ acts as internal time-parameter of the theory. Note that in the curvature squared theories handled earlier, the internal parameter was simply the proper volume $a^3$.

\subsection{Probability Interpretation and extremization of the effective potential:}

The continuity equation is

\be {dP\over d\sigma} + \mathbf{\nabla}.\mathbf{J} = 0,\ee
where, $P = \Psi^*\Psi$ is the probability density and $\mathbf{J}$ is the current density. Now in view of the quantum equation \eqref{Q2}, one can write,

\be \Psi^*{\partial\Psi\over \partial\sigma} = {i\hbar \Psi^* \over 198\gamma x(x^2 + 4k \sigma^{2\over 11})^2} \left({\partial^2\over \partial x^2} + {n\over x}{\partial\over \partial x}\right)\Psi -{i\over \hbar} V_e\Psi\Psi^*.\ee

\be \Psi{\partial\Psi^*\over \partial\sigma} = -{i\hbar\Psi \over 198\gamma x(x^2 + 4k \sigma^{2\over 11})^2} \left({\partial^2\over \partial x^2} + {n\over x}{\partial\over \partial x}\right)\Psi^* + {i\over \hbar} V_e\Psi\Psi^*.\ee
Therefore,

\bes \label{PI} {dP\over d\sigma} = \Psi^*{d\Psi\over d\sigma} + \Psi{d\Psi^*\over d\sigma} = {i\hbar (\Psi^*\Psi_{,xx} - \Psi\Psi^*_{,xx})\over 198\gamma x(x^2 + 4k \sigma^{2\over 11})^2} + {i\hbar n (\Psi^*\Psi_{,x} - \Psi\Psi^*_{,x})\over 198\gamma x^2(x^2 + 4k \sigma^{2\over 11})^2}\\&
= -{\partial\over \partial x}\Big[{i\hbar (\Psi\Psi^*_{,x} - \Psi^*\Psi_{,x})\over 198\gamma x(x^2 + 4k \sigma^{2\over 11})^2}\Big] - {i\hbar (\Psi\Psi^*_{,x} - \Psi^*\Psi_{,x})\over 198\gamma x^2(x^2 + 4k \sigma^{2\over 11})^2}\big[(n+5)x^2 + (n+1)4k \sigma^{2\over 11}\big].\\& = -{\partial J_x \over \partial x},
\end{split}\ee
provided $k = 0$ and $n = -5$, so that the second term on the right hand side of equation \eqref{PI} vanishes. Thus continuity equation holds only in the flat space, for $n = -5$, where, $\mathbf{J} = (J_x, 0, 0)$, and hence standard quantum mechanical probability interpretation holds in a straight forward manner. Note that in the flat space, the kinetic part does not show up explicit dependence on the internal time parameter $\sigma$, although the effective potential does. This makes the quantum equation easily tractable. Although, operator ordering ambiguity is somewhat fixed ($n = -5$) from physical consideration, viz. for probability interpretation to hold, nevertheless, the fact that it is different ($n = -1$) from the one obtained for $R^2$ theory of gravity, even with Gauss-Bonnet-Dilatonic coupling is a clear contradiction. The fact that probability interpretation holds in the flat space, i.e. for $k = 0$ may be interpreted in the following way. Either, Gauss-Bonnet squared term should not be included in the action, or, probabilistic interpretation holds only after the inflationary regime.\\

\noindent
The effective potential in the flat space reads as

\be\label{Ve} V_e(x,\sigma) = {2\over 11}\left[{3\alpha x \over 2\sigma^{10\over 11}} + {135\gamma x^7\over 28 \sigma^2}\right]  = {2\over 11}\left[{3\alpha x\over 2 z^5} + {135\gamma x^7\over 28 z^{11}}\right].\ee
The extremum of the effective potential with respect to $x$ then gives ($\sigma$ being the effective in-built time parameter of the theory),

\be {\partial V_e\over \partial x} = {2\over 11}\left[{3\alpha\over 2 z^5} + {135 \gamma\over 4}{x^6\over 4z^{11}} \right]=0 \Rightarrow {3\alpha\over 2}{x^2\over z^2} + {135\gamma\over 4}{x^8\over z^8} = 0.\ee
Thus, de-Sitter solution is obtained fixing the Lapse function $N = 1$, i.e. using the relation $\dot z = x$ as,

\be \label{z} z = z_0 \exp{\left[\left({2\alpha\over 45\gamma_1}\right)^{1\over 6}t\right]}  \Rightarrow a = a_0 e^{\lambda_1 t},~~ \mathrm{where}~~ \lambda_1 = {1\over 2}\left({2\alpha\over 45\gamma_1}\right)^{1\over 6} = {1\over 2}\left({M_P^2\over 45\gamma_1}\right)^{1\over 6},\ee
provided $\gamma_1 \rightarrow -\gamma_1$. The required choice that the coupling parameter has to be negative, of-course results in a clear contradiction with both the classical solution \eqref{Sol2} and the slow-roll inflationary solution \eqref{Inf}, which hold for positive coupling parameter $\gamma$. Thus we encounter the second pathology associated with the the theory under consideration.

\subsection{Semiclassical approximation:}

We can only perform semi-classical approximation on-shell in view of the classical solution \eqref{Sol2} or the Inflationary solution \eqref{Inf}. For this purpose, we express the wave-equation \eqref{Q1} in the following form ($k = 0,~ n = -5$)

\be\begin{split} \label{SC1} & -{\hbar^2 z^{9\over 2}\over 36\gamma x^5}\left({\partial^2\over \partial x^2} - {5\over x}{\partial\over \partial x}\right)\Psi - i\hbar {\partial \Psi\over \partial z} + \mathcal{V}(x, z)\Psi = 0\\&
\mathrm{where,}~ \mathcal{V}(x, z) = \left[{3\alpha x \over 2 \sqrt z} + {135\gamma x^7\over 28 z^{13\over 2}} \right].\end{split}\ee
The above equation may be treated as time independent Schr\"{o}dinger equation with two variables $x$ and $z$.
Hence as usual, let us seek the solution of the wave-function (\ref{SC1}) as,

\be\label{SC2} \Psi=\Psi_0 e^{\frac{i}{\hbar}S(x,z)}\ee
and expand $S$ in power series of $\hbar$ as,

\be\label{SC3} S = S_0(x, z) + \hbar S_1(x, z) + \hbar^2S_2(x, z) + .... .\ee
One can then find,

\bes\label{SC4}
\Psi_{,x} = {i\over \hbar}[S_{0,x}+\hbar S_{1,x}+\hbar^2 S_{2,x}+\mathcal{O}(\hbar)]\Psi;~~~~~\Psi_{,xx}
= {i\over \hbar}[S_{0,xx} + \hbar S_{1,xx} + \hbar^2 S_{2,xx}+ \mathcal{O}(\hbar)]\Psi\\&\hspace{0.34 in}
-{1\over \hbar ^2}[S_{0,x}^2 + \hbar^2 S_{1,x}^2 + \hbar^4 S_{2,x}^2 + 2\hbar S_{0,x} S_{1,x} + 2\hbar^2 S_{0,x} S_{2,x}
+ 2\hbar^3 S_{1,x} S_{2,x}+\mathcal{O}(\hbar)]\Psi;\\&\Psi_{,z} = {i\over \hbar}[S_{0,z}+\hbar S_{1,z}+\hbar^2 S_{2,z}
+\mathcal{O}(\hbar)]\Psi,\end{split}\ee
etc., where ``comma" in the suffix stands for derivative. Now, inserting $\Psi, \Psi_{,x}, \Psi_{,xx}, \Psi_{,z}$ etc. in view of \eqref{SC2}, and \eqref{SC4} in equation \eqref{SC1} and equating the coefficients
of different powers of $\hbar$ to zero, the following set of equations (upto second order) are obtained,

\begin{subequations}\begin{align}
&\label{SC5}\frac{z^{9\over 2}}{36\gamma x^5}S_{0,x}^2 + S_{0,z} + \mathcal{V} = 0\\
&\label{SC6}i\left[\frac{z^{9\over 2}S_{0,xx}}{36\gamma x^5} - \frac{5 z^{9\over 2} S_{0,x}}{36\gamma x^6}\right]-S_{1,z}
-\frac{z^{9\over 2} S_{0,x}S_{1,x}}{18\gamma x^5}=0\\
&i\left[\frac{z^{9\over 2} S_{1,xx}}{36\gamma x^5}-\frac{5 z^{9\over 2} S_{1,x}}{36\gamma x^6}\right]-S_{2,z} - \frac{z^{9\over 2}S_{0,x}S_{2,x}}{18\gamma x^5} - {z^{9\over 2} S_{1,x}^2\over 36\gamma x^5}=0
\end{align}\end{subequations}
which are to be solved successively to find $S_0(x, z)$, $S_1(x, z)$ and $S_2(x, z)$ and so on. Now, Identifying $S_{0,x}$ with $p_x$ and $S_{0,z}$ with $p_z$, the Hamilton constraint equation \eqref{H1} is retrieved. Further, using the definitions of momenta \eqref{He1} and equation (\ref{He2}) the $(^0_0)$ component of Einstein's equation \eqref{FE2} (for $N = 1, k =0)$ is also retrieved. So everything is consistent so far.\\

\noindent
Now, in view of the classical solutions \eqref{Sol2}, one can now compute the momenta \eqref{He2} and their integrals as

\be\label{ml}\begin{split}&
p_x = 18\gamma (2\lambda)^{11\over 2} \sqrt x;\hspace{1.0 in} p_z = -6\alpha\lambda\sqrt z - \frac{(2\lambda)^7\times 81\gamma\sqrt z}{7}\\&
\int p_x dx = (2\lambda)^{7}\times 12\gamma z^{3\over 2};\hspace{0.50 in}
\int p_z dz = -4\alpha\lambda z^{3\over 2} - \frac{(2\lambda)^7\times 54\times \gamma}{7} z^{3\over 2}
\end{split}\ee
Thus $S_{0}$, which when expressed in terms of the integrals of momenta yields

\be \label{HJ1} S_{0} = \int p_x dx + \int p_z dz = - 4\alpha\lambda z^{3\over 2} + \frac{3840}{7}\gamma\lambda^7 z^{3\over2}.\ee
One can also compute the zeroth order on-shell action \eqref{A1} in view of the classical solution \eqref{Sol2} as

\bes \label{HJ2} A_{0} = \int \left[-\frac{3\alpha \dot z^2}{2\sqrt z} + 24^2\gamma\left(\frac{\dot z^4\ddot z^2}{64 z^{9\over 2}}-\frac{15 \dot z^8}{1792 z^{13\over 2}}\right)\right] {dt}\\& = \int \left[-\frac{3\alpha (2\lambda)^2 z^2}{2\sqrt z} + 9\gamma\left(\frac{(2\lambda)^8 z^6}{z^{9\over 2}}-\frac{15 (2\lambda)^8 z^8}{28 z^{13\over 2}}\right)\right] {dz\over 2\lambda z} = - 2\alpha\lambda z^{3\over 2} + \frac{2496}{7}\gamma\lambda^7 z^{3\over2}.\end{split}\ee
Substituting $\gamma = {\alpha\over 96 \lambda^6}$ in view of relation \eqref{Sol2} both in \eqref{HJ1} and \eqref{HJ2}, one finally finds $A_0 = S_0 = {12\over 7} \alpha\lambda z^{3\over 2}$, and the classical on-shell action matches exactly with the Hamilton-Jacobi function. Alas! Equation \eqref{SC5} is not satisfied for the form of $S_0$ so obtained. Thus, we encounter yet another contradiction, the third pathology associated with the theory under consideration, and it is therefore useless to compute semiclassical wavefunction any further.

\section{Adding a cosmological constant}

In an attempt to find the phase-space structure corresponding to action \eqref{A1}, we have faced at least two pathologies, that has never been encountered with the higher-order gravitational actions handled so far. Firstly, the classical as well as the slow roll inflationary solutions require a positive coupling parameter $\gamma$, while the extremum of the effective potential requires negative. One can in no way expect that the classical solution should match the extremum of the potential. But appearance of the coupling parameter with opposite signs really matters. Next, and even more notorious problem that we encounter is the fact that the Hamilton-Jacobi function $S_0$ does not satisfy the Hamilton-Jacobi equation \eqref{SC5}. In order to overcome such pathologies, let us now modify the action \eqref{A1} taking cosmological constant ($\Lambda$) into account, which is essentially the vacuum energy density of all possible fields that exist in the very early universe. The modified action \eqref{A1} now reads as,

\be\label{A2} A = \int \left[ {R - 2\Lambda\over 16\pi G} + \beta \mathcal{G}^2\right]\sqrt{-g}~d^4 x,\ee
which in the R-W metric under consideration takes the following form

\be \label{A0}\begin{split} A = & \int\Bigg[6\alpha\left(-{\dot z^2\over 4 N \sqrt z}+ k N\sqrt z - {\Lambda\over 3}N z^{3\over 2}\right) + 24^2\gamma\Bigg\{\left({\dot z^4\over 64 N^7 z^{9\over 2}}+{k\dot z^2\over 8 N^5 z^{7\over 2}}+{k^2\over 4 N^3 z^{5\over 2}}\right)\ddot z^2\\&
-\left({\dot N \dot z^5\over 32 N^8 z^{9\over 2}} + {k\dot N\dot z^3\over 4 N^6 z^{7\over 2}} + {k^2\dot N\dot z\over 2 N^4 z^{5\over 2}}\right)\ddot z
-{15\dot z^8\over 1792 N^7 z^{13\over 2}}-{13 k\dot z^6\over 160N^5 z^{11\over 2}}-{11 k^2\dot z^4\over 48 N^3 z^{9\over 2}}\\&+{\dot N^2\dot z^6\over 64 N^9 z^{9\over 2}}+{k\dot N^2\dot z^4\over 8 N^7 z^{7\over 2}}+{k^2\dot N^2\dot z^2\over 4 N^5 z^{5\over 2}}\Bigg\} \Bigg]dt\int d^3x.\end{split}\ee
As before, in view of the new variable $\dot z = N x$, i.e. $\ddot z = N\dot x + \dot N x$, one can express the point Lagrangian in the form,

\be \label{L}\begin{split} L = &~ 6\alpha N\left(-{ x^2\over 4\sqrt z}+ k \sqrt z - {\Lambda\over 3} z^{3\over 2}\right) \\&
288\gamma \left[\left({x^4\over 32 N z^{9\over 2}}+{k x^2\over 4Nz^{7\over 2}}+{k^2\over 2Nz^{5\over 2}}\right)\dot x^2
- {15 N x^8\over 896 z^{13\over 2}} - {13 k N x^6\over 80 z^{11\over 2}} - {11 k^2 N x^4\over 24 z^{9\over 2}}\right],\end{split}\ee
after removing the total derivative terms from the action \eqref{A0}, following integration by parts. In the flat space $k = 0$, field equations read as,

\bes \label{fe1}2\alpha\left({\ddot z\over z}-{\dot z^2\over 4z^2} - \Lambda\right)+12\gamma\left[{\dot z^4\ddddot z\over z^5}+{8\dot z^3\ddot z\dddot z\over z^5}-{9\dot z^5\dddot z\over z^6}+{6\dot z^2\ddot z^3\over z^5}-{135 \dot z^4\ddot z^2 \over 4 z^6}+{159 \dot z^6\ddot z\over 4z^7} - {195 \dot z^8\over 16 z^8}\right] = 0.\end{split}\ee

\be \label{fe2} 2\alpha \left({3\dot z^2\over 4z^2}-\Lambda\right) + 18\gamma\left[{\dot z^5 \dddot z\over z^6} + {3\dot z^4 \ddot z^2\over 2z^6} - {9 \dot z^6\ddot z\over 2 z^{7}} + {15 \dot z^8\over 8 z^8}\right] = 0.\ee
The above field equations admit the following exponential solution in vacuum

\be\label{sol} z = z_0 e^{2\lambda t},~~\mathrm{under~the~condition}~~~ \Lambda = 3\lambda^2 - 288{\gamma\over\alpha} \lambda^8.\ee
To keep $\Lambda > 0$, the condition required is $\alpha > 96\gamma \lambda ^6$. One can check that the solution reduces to \eqref{Sol2} setting $\Lambda = 0$. One can further, perform slow-roll approximation, which finally ends up with

\be 96\gamma H^8 - \alpha H^2 - {4\over 3} \alpha \Lambda =0. \ee
The above algebraic equation equation for the Hubble parameter $H$, may be solved to obtain eight ($8$) real roots, and as a result the scale factor admits exponential expansion. We do not present the solutions to avoid unnecessary complications. We can now perform the constraint analysis as before, to end up with the following Hamiltonian,

\be\label{H2}\ H = N\left[x p_z + {z^{9\over 2}p_x^2\over 36\gamma(x^2 + 4 k z)^2} +\alpha\left({3 x^2\over 2\sqrt z}- 6k \sqrt z +2\Lambda z^{3\over 2}\right)+36\gamma x^4\left({15 x^4\over 112 z^{13\over 2}} + {13 k x^2\over 10 z^{11\over 2}} + {11 k^2\over 3 z^{9\over 2}}\right)\right] = N\mathcal{H}.\ee

\noindent
The action \eqref{A0} may now be expressed in canonical ADM form ($k = 0$) as,

\be A = \int\big(\dot z p_z + \dot x p_x  - \ N \mathcal{H}\big) dt d^3 x = \int\big(\dot {h}_{ij}p^{ij} + \dot {K}_{ij}\Pi^{ij} - \ N \mathcal{H}\big) dt d^3 x.\ee

\noindent
One can follow the same procedure towards canonical quantization of the Hamiltonian \eqref{H2}. It does not make any considerable change in the quantum equation \eqref{Q2}, rather it just modifies the effective potential appearing in equation \eqref{SC1} following an additive term involving cosmological constant $\Lambda$ to,

\bes \label{Q3} V_e(x,\sigma) =
\left[{3\alpha\over 11 x} \left({x^2\over \sigma^{10\over 11}}- {4k\over \sigma^{8\over 11}} + {4\Lambda\over 3 \sigma^{6\over 11}}\right) + {72\gamma x^3\over 11}\left({15 x^4\over 112 \sigma^2} + {13 kx^2\over 10 \sigma^{20\over 11}} + {11 k^2\over 3 \sigma^{8\over 11}} \right) \right].\end{split}\ee
As a result the probability interpretation remains unaltered, i.e. it holds in flat space $k = 0$, for operator ordering index $n = -5$. Now, extremization of the effective potential ($k = 0$) leads to

\be \label{z1} z = z_0 e^{2\lambda t},~~\mathrm{or~equivalently~} a = a_0 e^{\lambda t},~~\mathrm{for}~~\Lambda = 3 \lambda^2 + 4320{\gamma\over \alpha} \lambda^8.\ee
Of-course, the classical solution \eqref{sol} differs from \eqref{z1} to the extent of the condition required to impose on the cosmological constant. Nonetheless, classical solution is not expected to match at the extremum of the potential and therefore has nothing to do with consistency. Whatsoever, the important development is, it does not require a reverse sign of $\gamma$ unlike the situation encountered in solution \eqref{z} without cosmological constant. One can again check that the above condition \eqref{z1} reduces to the previous one \eqref{z} setting $\Lambda = 0$. Thus the pathology is removed, and cosmological constant saves the soul.\\

\noindent
In the context of semiclassical approximation, we express the quantized equation in the form \eqref{SC1} just with the modified effective potential,

\be\mathcal{V_e} (x, z) = \left[{3\alpha x \over 2 \sqrt z} + {135\gamma x^7\over 28 z^{13\over 2}} + 2\alpha \Lambda {z^{3\over 2}\over  x}\right],\ee
and proceed as before. The momenta remains unaltered and one can thus calculate the Hamilton-Jacobi function as,

\be S_0 = - 4\alpha\lambda z^{3\over 2} + {3840\over 7}\gamma \lambda^7 z^{3\over 2},\ee
which is the same as \eqref{HJ1}. However, zeroth order classical on-shell action is calculated from \eqref{A0} in the following manner.

\bes A_0 = \int \left[-6\alpha\lambda^2 z^{3\over 2} - 6\alpha \lambda^2 z^{3\over 2} + 576\gamma \lambda^8 z^{3\over 2} + 9\gamma (16^2 -{15\times 256\over 28})\lambda^8 z^{3\over 2}\right]dt\\&
= \int\left[-12\alpha \lambda^2 z^{3\over 2} + {11520\over 7} \gamma\lambda^8 z^{3\over 2}\right]{dz\over 2\lambda z}~~~
= \int\left[-6\alpha \lambda \sqrt z + {5760\over 7}\gamma\lambda^7 \sqrt z\right]dz\\&
= -4\alpha\lambda z^{3\over 2} + {3840\over 7}\gamma\lambda^7 z^{3\over 2}.\end{split}\ee
Since, Hamilton-Jacobi equation matches zeroth order on-shell action, so this part is well-behaved. One can check that the Hamilton-Jacobi function $S_0$ now satisfies the Hamilton-Jacobi equation \eqref{SC5} under the condition,

\be {\alpha^2\over 128\gamma\lambda^5} + {63648\over49}\gamma\lambda^7 - {15\over 7}\alpha\lambda = 0,\ee
which simply restricts $\lambda$ by the other two parameters ($\alpha$ and $\gamma$) of the theory. Therefore as mentioned Every thing is now consistent under the addition of a cosmological constant. The semiclassical wavefunction upto first order approximation now reads as,

\be \Psi = \Psi_0 e^{\left(-4\alpha \lambda + {3840\over 7}\gamma \lambda^7\right)z^{3\over 2}}.\ee
In order to compute the wavefunction upto first order of approximation, we note that in view of classical solution \eqref{Sol2}, the independent variables $x$ and $z$ are related as, $x = 2\lambda z$. Thus one can compute $S_{0,x},~ S_{0,xx}$ and also express $S_{1,x}$ in terms of $S_{1,x}$ to finally obtain,

\be S_{1,z} = - i \left(\frac{63}{28}\right)\left(\frac{7\alpha - 960\gamma \lambda^6}{7\alpha - 3648 \gamma \lambda^6}\right){1\over z},\ee
which may be integrated to obtain
\be S_1 = - i \left(\frac{63}{28}\right)\left(\frac{7\alpha - 960\gamma \lambda^6}{7\alpha - 3648 \gamma \lambda^6}\right) \ln z,\ee
apart from a constant of integration. Hence, the wave function upto first order of approximation finally reads as,

\be \label{SCW1}\Psi = \Psi_0 e^{\left(-4\alpha \lambda + {3840\over 7}\gamma \lambda^7\right)z^{3\over 2}}\times \left[z^{\left(\frac{63}{28}\right)\left(\frac{7\alpha - 960\gamma \lambda^6}{7\alpha - 3648 \gamma \lambda^6}\right)}\right].\ee
In this manner, it is possible to find the semiclassical wave function for even higher order of approximation. Since comparison with the classical Hamiltonian constraint equation reveals $p_x = {\partial S_0\over \partial x}$, and $p_z = {\partial S_0\over \partial z}$, so the wavefunction shows a strong correlation between coordinates and momenta. Now using the relation between velocities and momenta and the fact that $S_0$ obeys Hamilton-Jacobi equation, it is apparent that the above relations define a set of trajectories in the $x - z$ plane, which are solutions to the classical field equations. Thus the semiclassical wave function \eqref{SCW1} is strongly peaked around classical inflationary solutions \eqref{sol}. \\

\section{Conclusion}

Modified theory of gravity has been envisaged as an alternative to quintessence, being able to unify early inflation with late time cosmological evolution. Gauss-Bonnet-dilatonic coupled action is also a candidate in this regard. However, in the absence of a scalar field in the late stage of cosmic evolution, higher powers of Gauss-Bonnet term serves the purpose as well. It is therefore required to study the behaviour of a theory that modifies the Einstein-Hilbert action in the presence of Gauss-Bonnet squared term, at the first place. A theory with $f(R, \mathcal{G})$ gravity has also been found to unify early inflation with late stage of cosmic acceleration. We therefore, formulated the phase-space structure of such an action and followed standard canonical quantization scheme. In the process, the action has been found to suffer from a couple of pathologies. Firstly, the theory admits de-Sitter solution in vacuum and also exhibits exponential expansion under slow roll condition. Nevertheless, although it exhibits the same feature from the extremum of the effective potential, the coupling parameter $\gamma$ appears with reverse sign. Next, the form of the Hamilton-Jacobi function obtained under semiclassical approximation doesn't satisfy the Hamilton-Jacobi equation. We have therefore improvised the action under the addition of a cosmological constant term to observe that such pathologies are removed, leading to mathematical consistency of the theory. This proves the very importance of considering the presence of cosmological constant, which is essentially the sum of zero point energies of all quantum fields, available in the very early universe. On the other way round, the creation of the universe was initiated with non-trivial vacuum. The other pathology is non-appearance of a power law solution of the scale factor in the radiation dominated era. In fact, with the seeds of perturbation developed in the Inflationary regime, a Friedmann-like radiation dominated era ($a(t) \propto \sqrt t$) exactly can formulate the structures of the universe, we presently observe. Further, the CMBR is also an artefact of Friedmann-like radiation dominated era. We think that this pathology may only be circumvented if the action is further modified by the inclusion of additional curvature scalars or some form of matter fields.\\

\noindent
Acknowledgment: We would like to thank Prof. Salvatore Capozziello of Dipartimento di Fisica, Universita di Napoli, Federico II, Italy, for bringing our attention to this particular problem.

\end{document}